\documentclass[a4paper,
              ]{jacow}
\makeatletter%
	\ifboolexpr{bool{xetex}}
	 {\renewcommand{\Gin@extensions}{.pdf,%
	                    .png,.jpg,.bmp,.pict,.tif,.psd,.mac,.sga,.tga,.gif,%
	                    .eps,.ps,%
	                    }}{}
\makeatother

\ifboolexpr{bool{xetex} or bool{luatex}} 
 {}                                      
 {\usepackage[utf8]{inputenc}}           

\usepackage[final]{pdfpages}
\usepackage{multirow}
\usepackage{ragged2e}

\ifboolexpr{bool{jacowbiblatex}}%
 {%
  \addbibresource{jacow-test.bib}
  \addbibresource{biblatex-examples.bib}
 }{}
\begin{document}

\title{A New Damper for Coupled-Bunch Instabilities caused by the accelerating mode at \NoCaseChange{SuperKEKB}}

\author{Kouki Hirosawa\thanks{hirosawa@post.kek.jp}, 
		Kazunori Akai\textsuperscript{1}, Eizi Ezura\textsuperscript{1}, Tetsuya Kobayashi\textsuperscript{1}, \\Kota Nakanishi\textsuperscript{1}, Shin-ichi Yoshimoto\textsuperscript{1}, \\
		SOKENDAI (the Graduate University for Advanced Studies), Tsukuba, Japan,\\
		\textsuperscript{1}also at High Energy Accelerator Research Organization (KEK), Tsukuba, Japan}
	
\maketitle

\footnote[0]{The authors of this work grant the arXiv.org and LLRF Workshop's International Organizing Committee a non-exclusive and irrevocable license to distribute the article, and certify that they have the right to grant this license.}

\begin{abstract}
	SuperKEKB is an asymmetric electron-positron circular collider based on nano-beam scheme at interaction region and large beam current. Large beam current makes growth rates of longitudinal coupled-bunch instabilities (LCBI) large. Especially some lowest modes near accelerating frequency are serious. On the design parameter for SuperKEKB, $\mu$ = -1, -2, -3 mode of LCBI can be destabilized. We developed new LCBI damper to suppress newly arisen LCBI modes ($\mu$ = -1, -2, -3) in SuperKEKB. The new damper will be installed in Low Level RF control system. The new LCBI damper is independent of main LLRF control components.  In the test bench measurement, our new LCBI damper has good performance and satisfied required specifications. For preparation of using LCBI damper, we produced the simulation of beam oscillation damped by RF feedback. The results of this simulation shows that we need more dampers than them we prepared. We report profile of new damper and results of test bench measurement and feedback simulation in this paper.
\end{abstract}

\section{Introduction}
	SuperKEKB\cite{1} is an asymmetric energy electron-positron collider upgraded from KEKB. The SuperKEKB Phase-1 commissioning was operated from February to June in 2016, and the Phase-2 will be carried out from January in 2018.
	The aim of this accelerator is to find new physics beyond the Standard Model from point of view of event probability. For increasing event frequency, we are making accelerator higher luminosity by large beam current and the scheme of nano-size beam at interaction point.\\
	\hspace{1em}As the beam current increase, we have to detune the cavity resonant frequency. Cavity detuning causes coupled bunch instabilities modes around accelerating frequency. 
	Although we use low detuning normal conducting cavity, Some lowest modes of LCBI can be destabilized.
This instability give a limit of beam current without suppressing. A kind of this suppressing component has been already existed for KEKB operation\cite{2}, but it has less performance for using in SuperKEKB. So we need new damper which has enough specification to stabilize newly excited LCBI. The SuperKEKB storage ring consists of a 7 GeV high-energy ring (HER) for electrons and a 4 GeV low-energy ring (LER) for positrons. The exact value of growth rate for the each ring is different, but both of them will have first ($\mu = -1$) and second ($\mu = -2$) lowest mode of LCBI on the design parameter. In test bench measurement, our new damper gave quite good performance, and we could confirm the result of improvement for each components of LCBI damper\cite{3}. Table 1 provides the main parameters of SuperKEKB\cite{1} for this study. Below results of calculations follow this parameters.
\begin{table}[!htb]
		\centering
			\caption{SuperKEKB and Cavity Parameter}
			\begin{tabular}{l|cc} \hline
				parameters & \multicolumn{2}{c}{value} \\ \hline
				for SuperKEKB & LER & HER \\ \hline
				$Energy :E$ & 4.0GeV & 7.0GeV\\
				$Beam\ current :I_0$ & 3.6A & 2.62A \\
				$Mom.\ compact. :\alpha_c$ & $3.25 \times 10^{-4}$ & $4.55 \times 10^{-4}$\\
				$Synch.\ freq. :f_s$ & $2.43 kHz$ & $2.78 kHz$\\
				$Harmonic \ number :h$ &  \multicolumn{2}{c}{5120}\\ 
				$RF\ frequency :f_{rf}$ & \multicolumn{2}{c}{$508.877$ MHz}\\
				$f_{0} = f_{rf} / h$ & \multicolumn{2}{c}{$99.39$ kHz}\\
				Number of cavity & 22 & ARES 8, SC 8 \\ \hline
				for Cavity & ARES & SC \\ \hline
				$V_c / cavity$ & $0.5$MV &$1.5$MV\\
				$R_{s}/Q_0$ & 15$\Omega$ & 93$\Omega$\\
				$Q_0$ & $1.1 \times 10^5$ & $2.0 \times 10^9$ \\
				coupling factor :$\beta$ & $5.0$ & $4.0 \times 10^4$\\
				\hline
			\end{tabular}
	\end{table}

\section{Coupled-Bunch Instabilities Caused by Accelerating Mode}
\begin{figure}[!htb]
		\centering
		\includegraphics*[width=\linewidth, bb = 0 0 472 132]{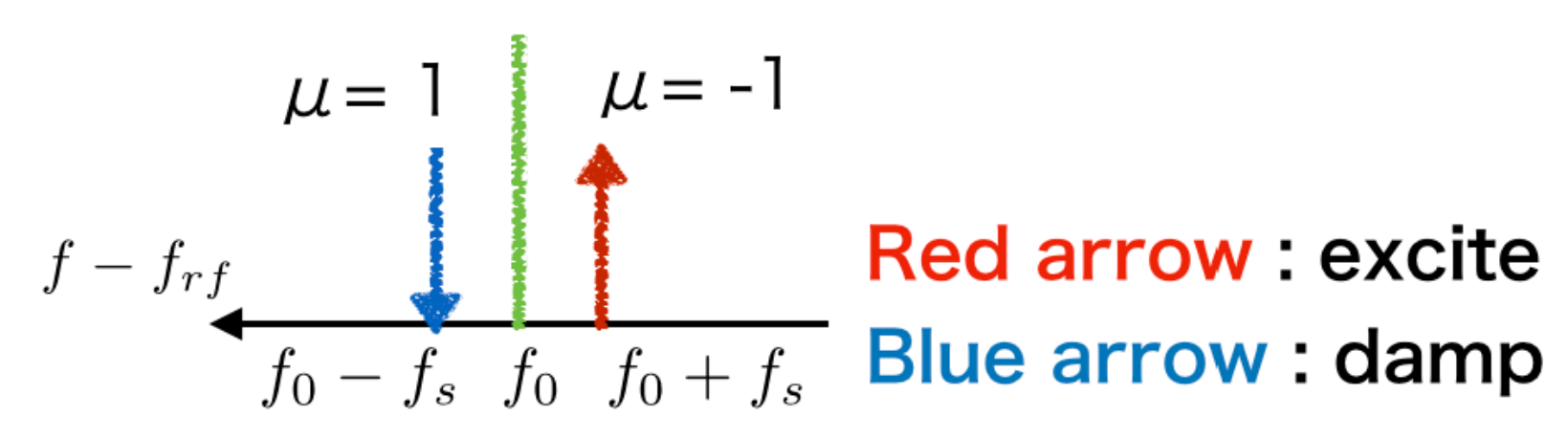}\\
		\includegraphics*[width=\linewidth, bb = 0 0 403 281]{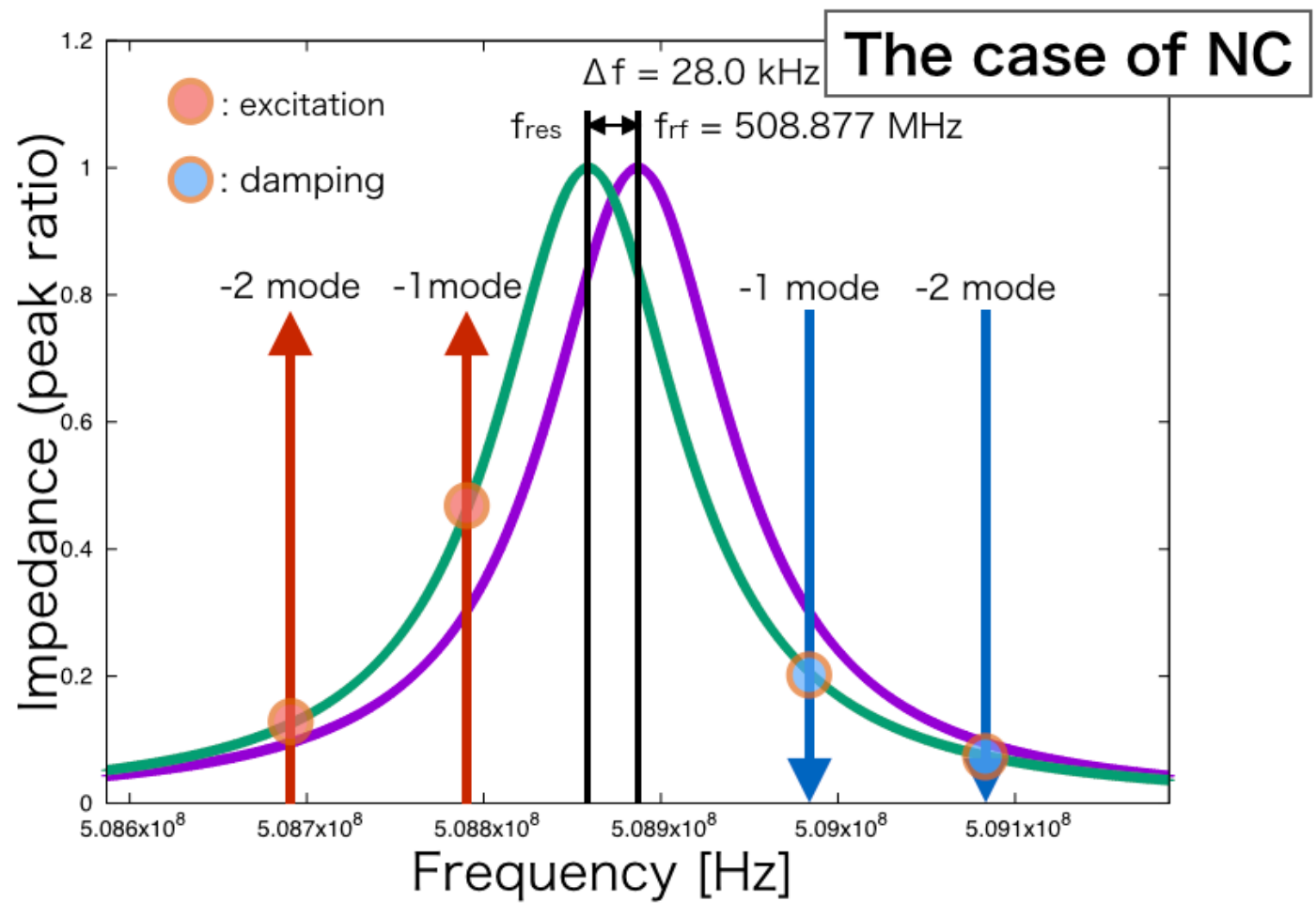}
		\caption{The positional relations between exiting term and damping term for LCBI growth rates(Top), and the cavity impedance and LCBI modes(Bottom). }
		\label{fig:0}
	\end{figure}

\begin{figure}[!htb]
		\centering
		\includegraphics*[width=\linewidth, bb = 0 0 753 534]{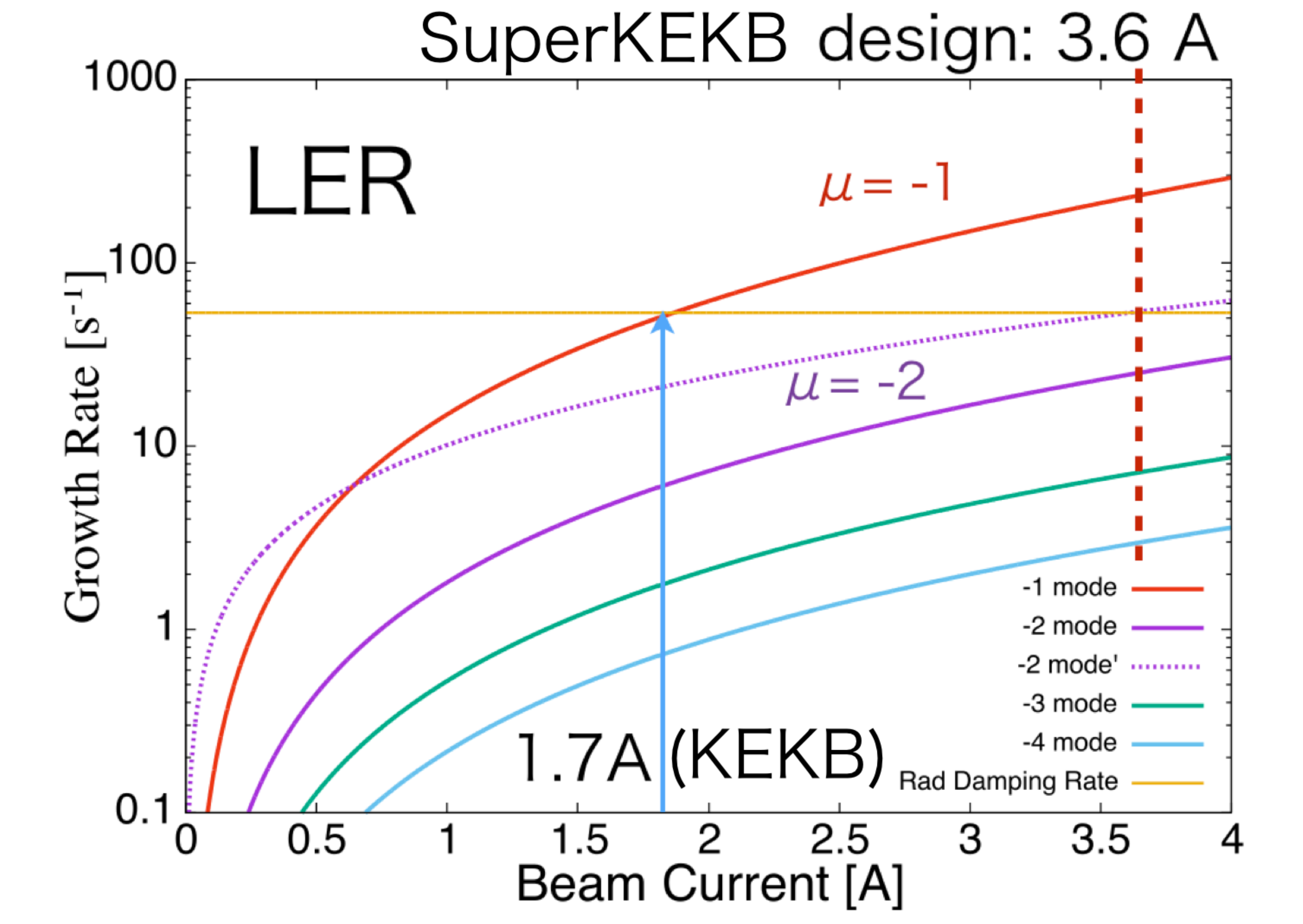}\\
		\includegraphics*[width=\linewidth, bb = 0 0 732 555]{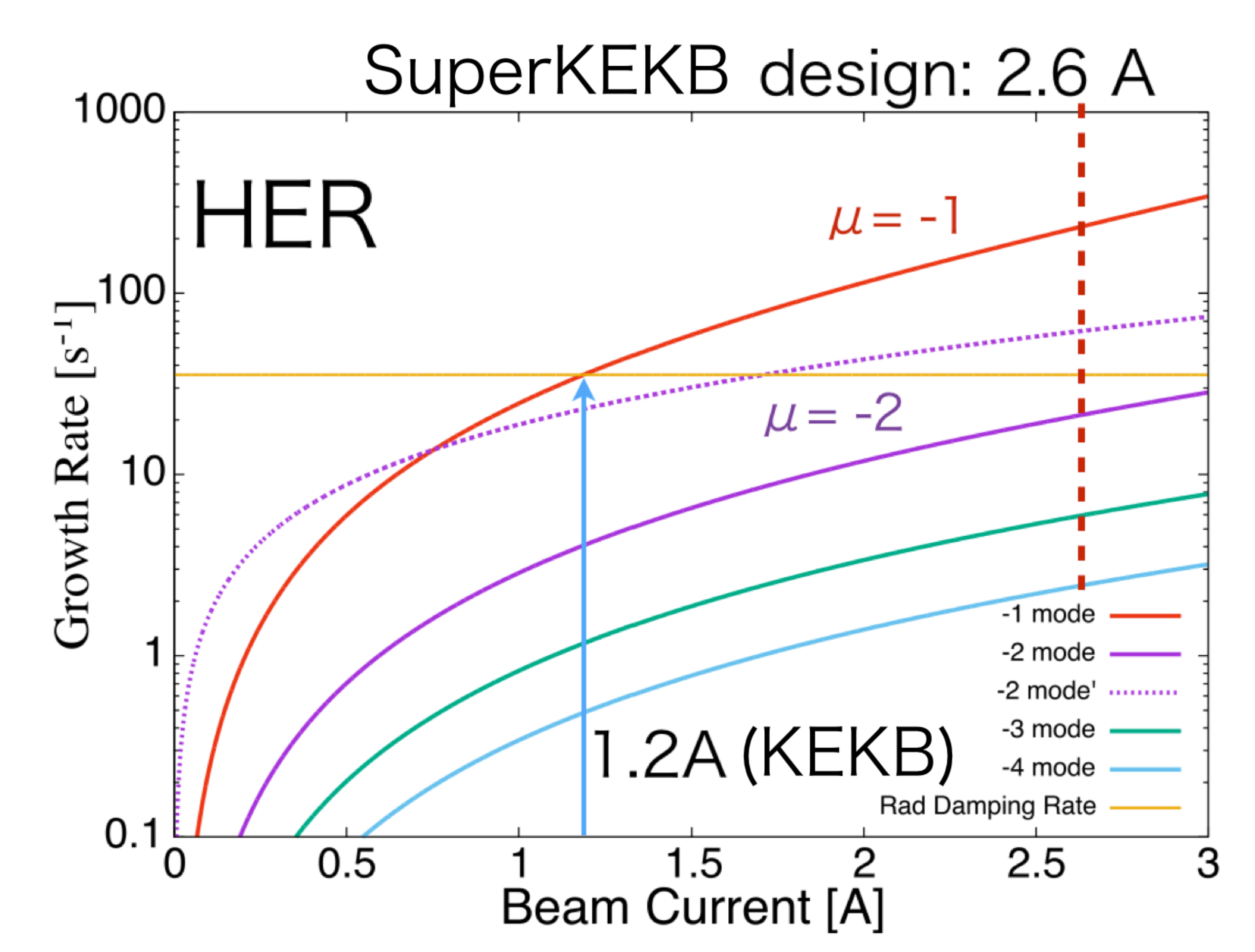}
		\caption{Growth rate by beam current. Dotted curve indicates $\mu$ = -2 mode in the case of a cavity parked with 150-kHz detuning.}
		\label{fig:1}
	\end{figure}

SuperKEKB has two types of cavities: superconducting and normal conducting cavities\cite{4}. In HER, a normal conducting cavity (NC) and superconducting cavity (SC) are used, and in LER, only NC. NC has a three-cavity structure which consists of accelerating, energy storage, and coupling cavity. Thanks for the energy storage cavity, we can reduce detuning value. This NC which has unique structure for KEKB/SuperKEKB is called "ARES"\cite{5}.
Nevertheless this structure makes detuning value small for heavy beam loading, $\mu = -1, -2, -3$ modes can be destabilized on SuperKEKB design parameter. 
The superconducting cavity has typical one-cell structure, and they will be set near detuning amount to ARES.
So our LCBI problem for accelerating mode is enhanced at $\mu = -1$ and $\mu = -2$ mode. 
Several mode of LCBI are growing as below equation.
	\begin{align}
		\tau_{\mu}^{-1} &= AI_0
		 \sum^\infty_{p= 0} \{ f^{(\mu+)}_p \mathrm{Re} Z^{||}(f^{(\mu+)}_p)
		- f^{(\mu-)}_p \mathrm{Re} Z^{||}(f^{(\mu-)}_p) \}\\
		f^{(\mu+)}_p &= (p M+ \mu)f_0+ f_s, \ \ \ 
		f^{(\mu-)}_p = \{(p+ 1)M- \mu\} f_0- f_s \nonumber\\
		f_{rf} &=h f_0 \ \ \ \ ,\nonumber
	\end{align}
	where $\tau_{\mu}^{-1}$ is growth rate, $Z^{||}$ is longitudinal impedance, $f_s$ is synchrotron frequency, and $h$ is harmonic number. Detuning impedance condition was applied to resonance frequency of $Z^{||}$.
	\\ \hspace{1em}
	The positional relations between the cavity impedance and LCBI modes are illustrated in Figure \ref{fig:0}.
	The top of Figure \ref{fig:0} explains the relation of positions for small span around $f_{rf} - f_0$ ($f_{rf}$ is accelerating frequency and $f_0$ is revolution frequency). Red arrow denotes exciting term of growth rate and Blue term denotes damping term of growth rate. A little bit difference between exciting and damping terms has great importance in this damper.
	The bottom of Figure \ref{fig:0} explains illustration around accelerating frequency for the cavity impedance of ARES.
	Figure \ref{fig:1} shows calculations of growth-rate for LCBI modes caused by accelerating mode (Top : LER, Bottom : HER).
	Solid curves are growth rate of several modes on the ideal parameters.
	Dotted curves are growth rate under the special case that we have one parked cavity on manual detuning by RF system trouble. The parked cavity is ARES for both rings. Horizontal yellow line in the figure indicates the radiation damping rate. From this calculation, suppression of $\mu = -1$ and $-2$ modes must be required for the design beam current. Furthermore the $\mu = -3$ mode LCBI should be concerned for the new damper development in preparation for unexpected problems.
	
\section{The system of new LCBI Damper}
\begin{figure}[!htb]
		\centering
		\includegraphics*[width=\linewidth, bb = 0 0 854 228]{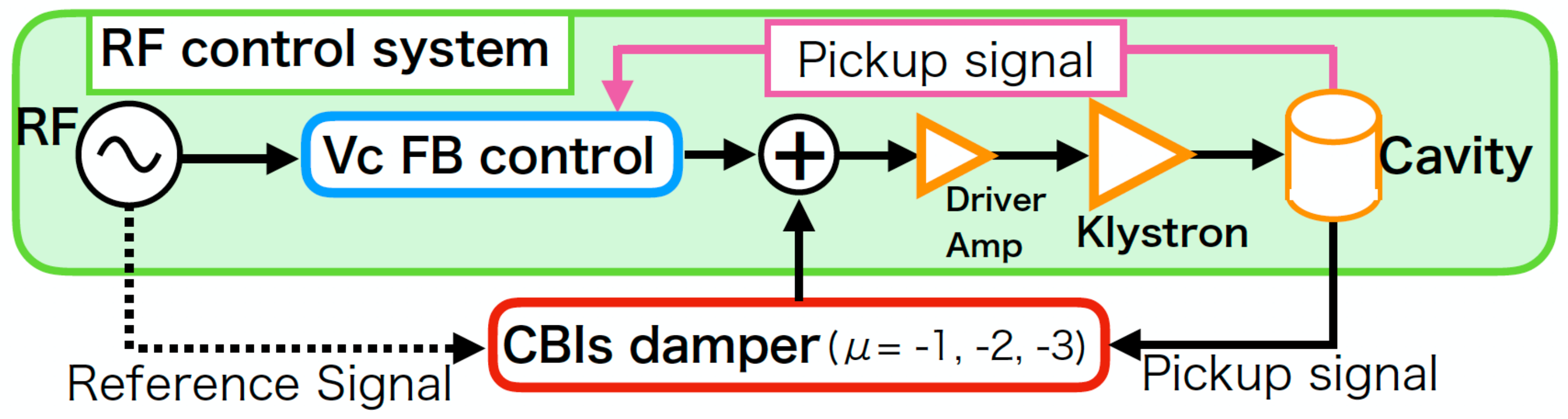}\\
		\caption{Block diagram of an RF FB system for LCBI damper.}
		\label{fig:3}
	\end{figure}

\begin{figure}[!htb]
		\centering
		\includegraphics*[width=\linewidth, bb = 0 0 758 327]{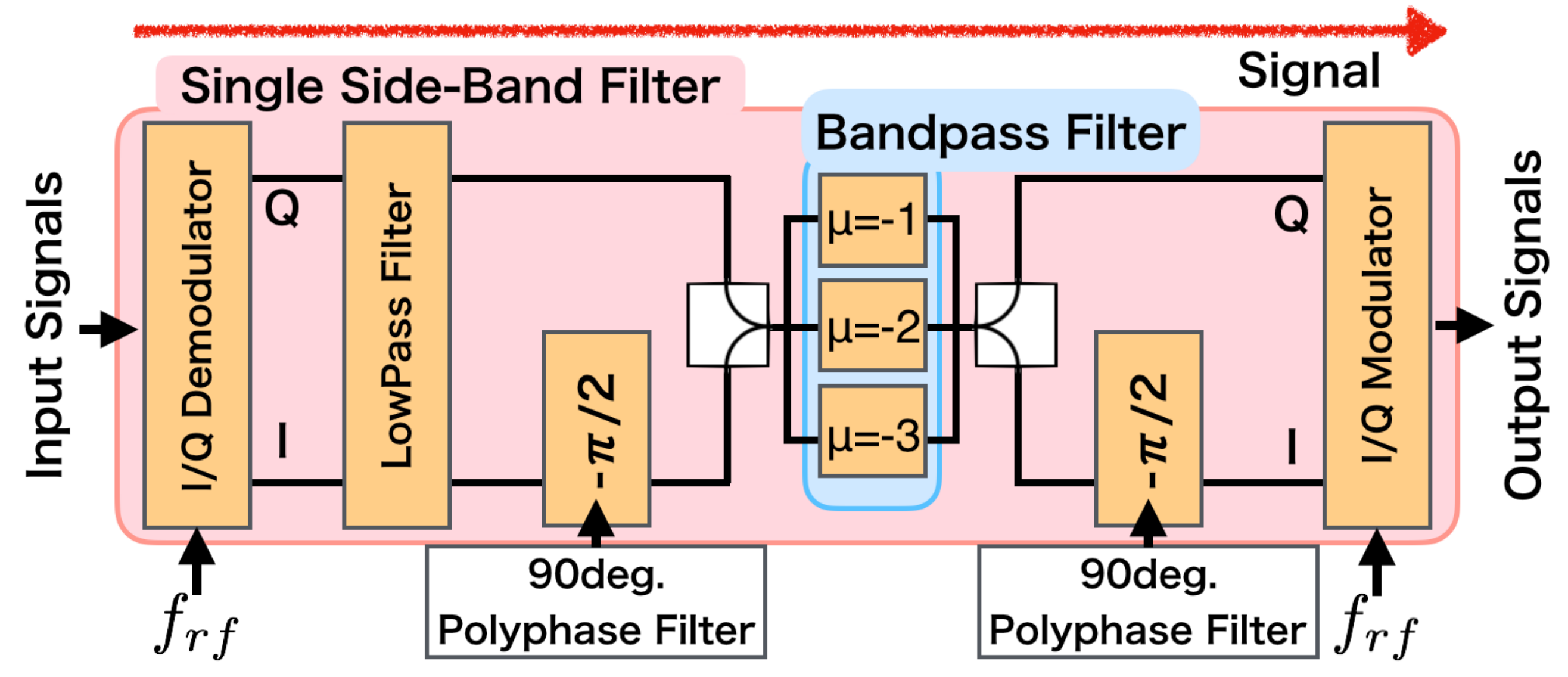}\\
		\includegraphics*[width=\linewidth, bb = 0 0 683 242]{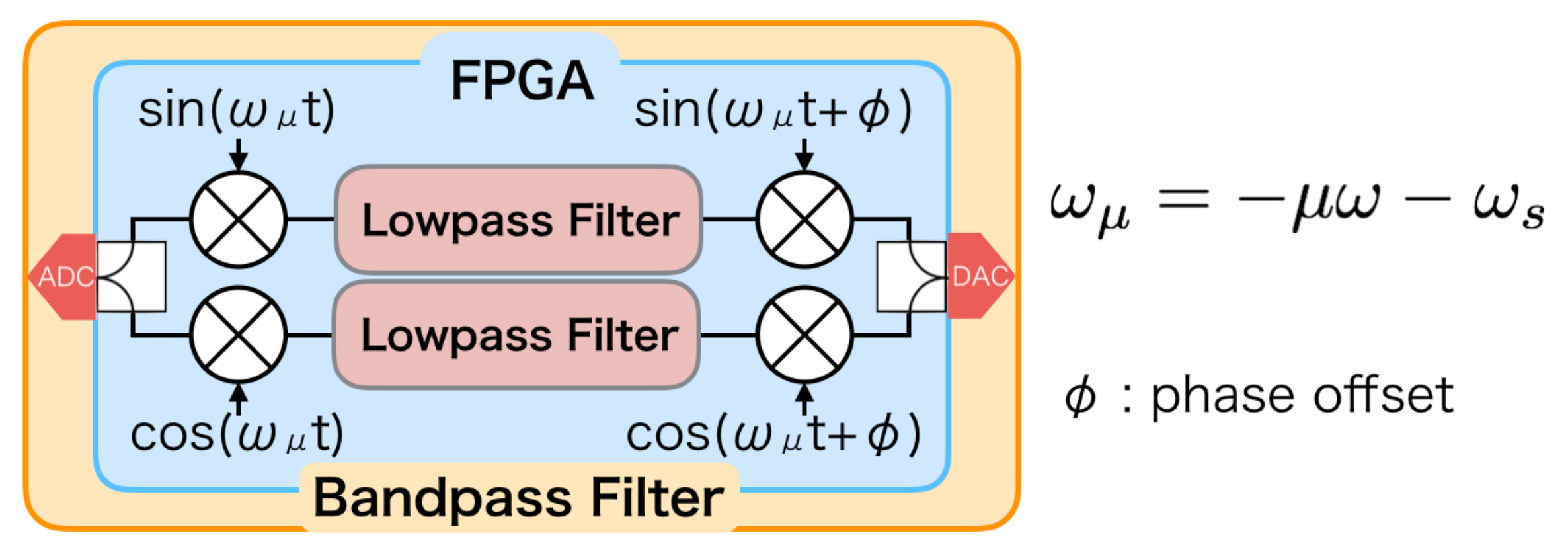}
		\caption{Block diagram of a new LCBI damper. Upper figure is entire structure of LCBI damper, and lower figure is internal structure of BPF.}
		\label{fig:4}
	\end{figure}

\begin{figure}[!htb]
		\centering
		\includegraphics*[width=\linewidth, bb = 0 0 517 419]{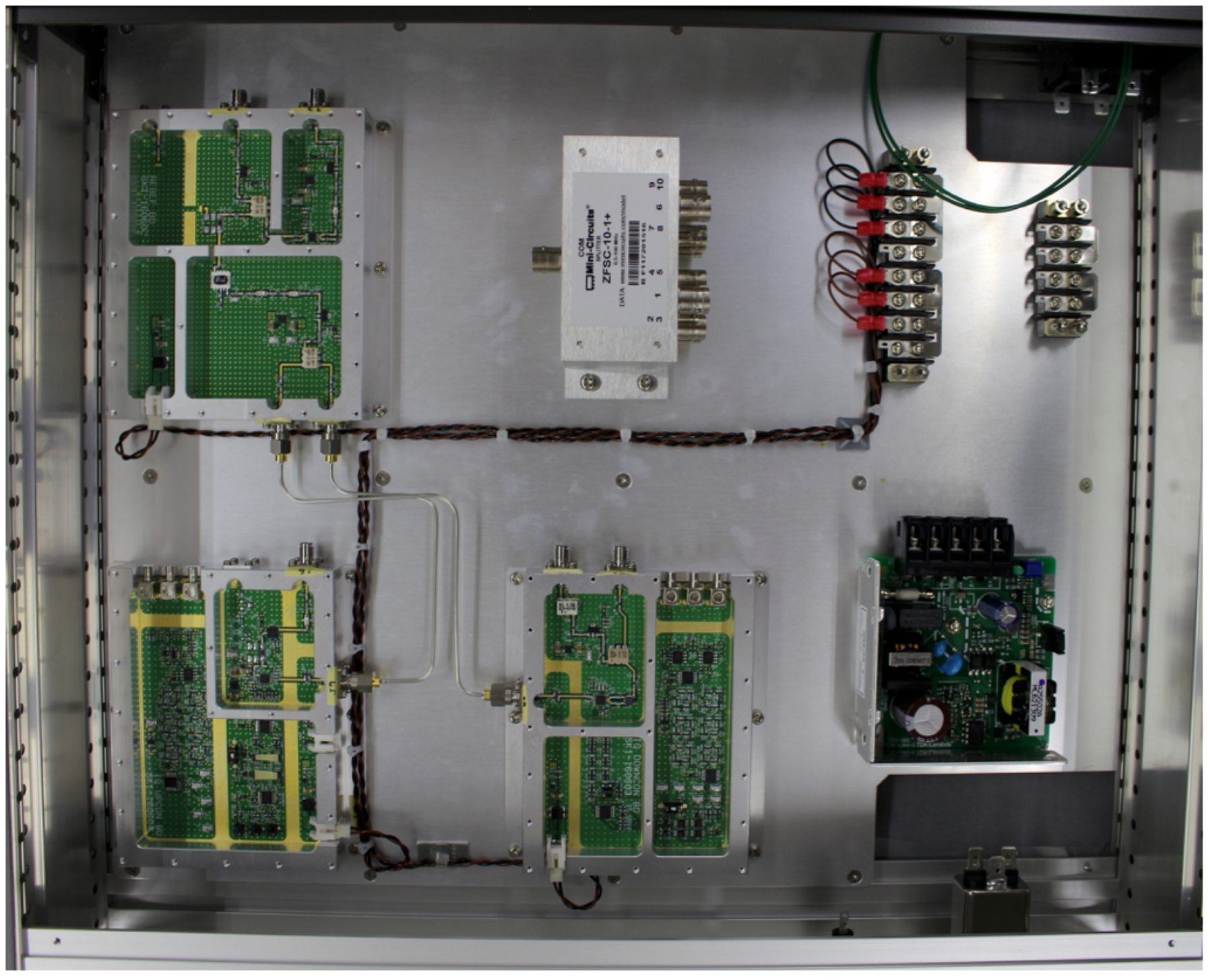}\\
		\includegraphics*[width=\linewidth, bb = 0 0 596 524]{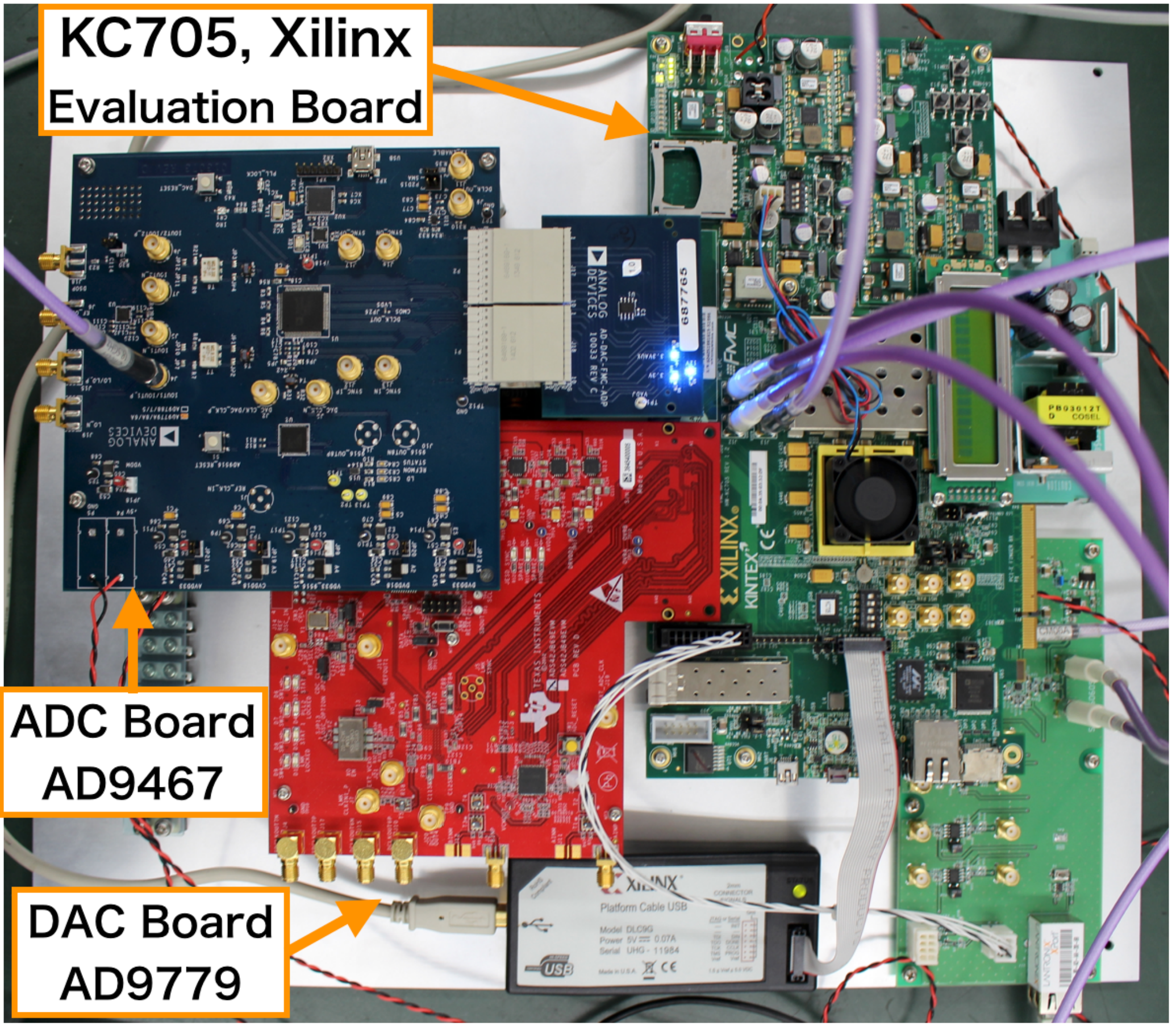}\\
		\caption{Pictures of LCBI dampers. The top figure is internal circuit of SSBF, and the bottom figure is BPF.}
		\label{fig:5}
	\end{figure}

	To suppress the synchrotron oscillation for LCBI destabilized modes, we will use LCBI damper as an RF feedback (FB).
	LCBI damper signals are combined to Vc FB control signals.
	Some LLRF control systems including Vc FB are digitalized by FPGA and $\mu$TCA\cite{6}.
	LCBI damper is connected with this LLRF control system for actual use.
	Figure \ref{fig:3} shows a block diagram of the FB system for LCBI damping.
	\\ \hspace{1em}
	The LCBI damper consists of single-side band filter (SSBF) which is analog circuit and digital bandpass filter (BPF) which is made on FPGA board (KC705, Xilinx)\cite{3}.
	The fundamental method of this system is based on the KEKB damper.
	First step of LCBI damper, only lower sideband of transmit signals can be passed through SSBF.  RF are input as reference signals for up/down conversion.
	Second step, baseband signals are filtered by bandpass filter (BPF), and then remaining signals are only LCBI $\mu = -1, -2, -3$ modes (Fig. \ref{fig:4} bottom).
	To suppress LCBI $\mu = -1, -2, -3$ modes at the same path of a RF system, three digital BPFs are combined to a SSBF in parallel (Fig. \ref{fig:4} top).
	\\ \hspace{1em}
	Our klystron has very narrow bandwidth ($<300kHz$), and we have to adjust phase of LCBI modes to damp only exciting effects and to keep damping effects as much as possible.
	So we don't use comblike filter by one turn delay for bandpass filtering of each LCBI modes because phase controlling is difficult for comblike filter.
	Since BPFs can be adjusted the phase parameter independently for each LCBI mode, feedback signals are optimized conforming with the phase characteristics of several components in entire RF system.
	The control parameters of the digital BPF can be given remotely by EPICS records.
	Figure \ref{fig:5} shows pictures of LCBI dampers.
	
	\section{Performance of New Damper}
\begin{figure}[!htb]
		\centering
		\includegraphics*[width=\linewidth, bb = 0 0 569 444]{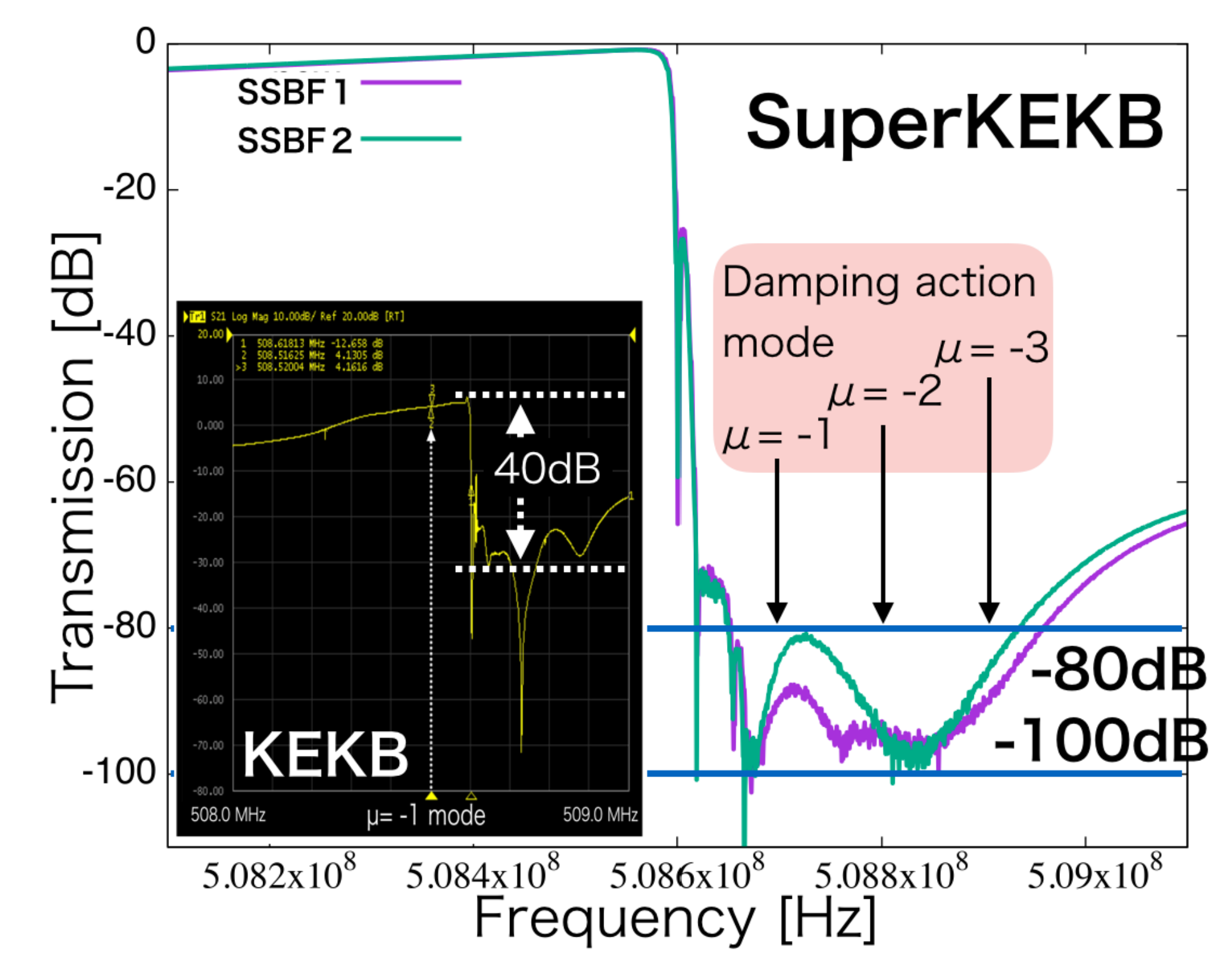}
		\caption{Frequency characteristics of SSBF transmission and that of KEKB damper.}
		\label{fig:6}
	\end{figure}

\begin{figure}[!htb]
		\centering
		\includegraphics[width=\linewidth, bb = 0 0 916 610]{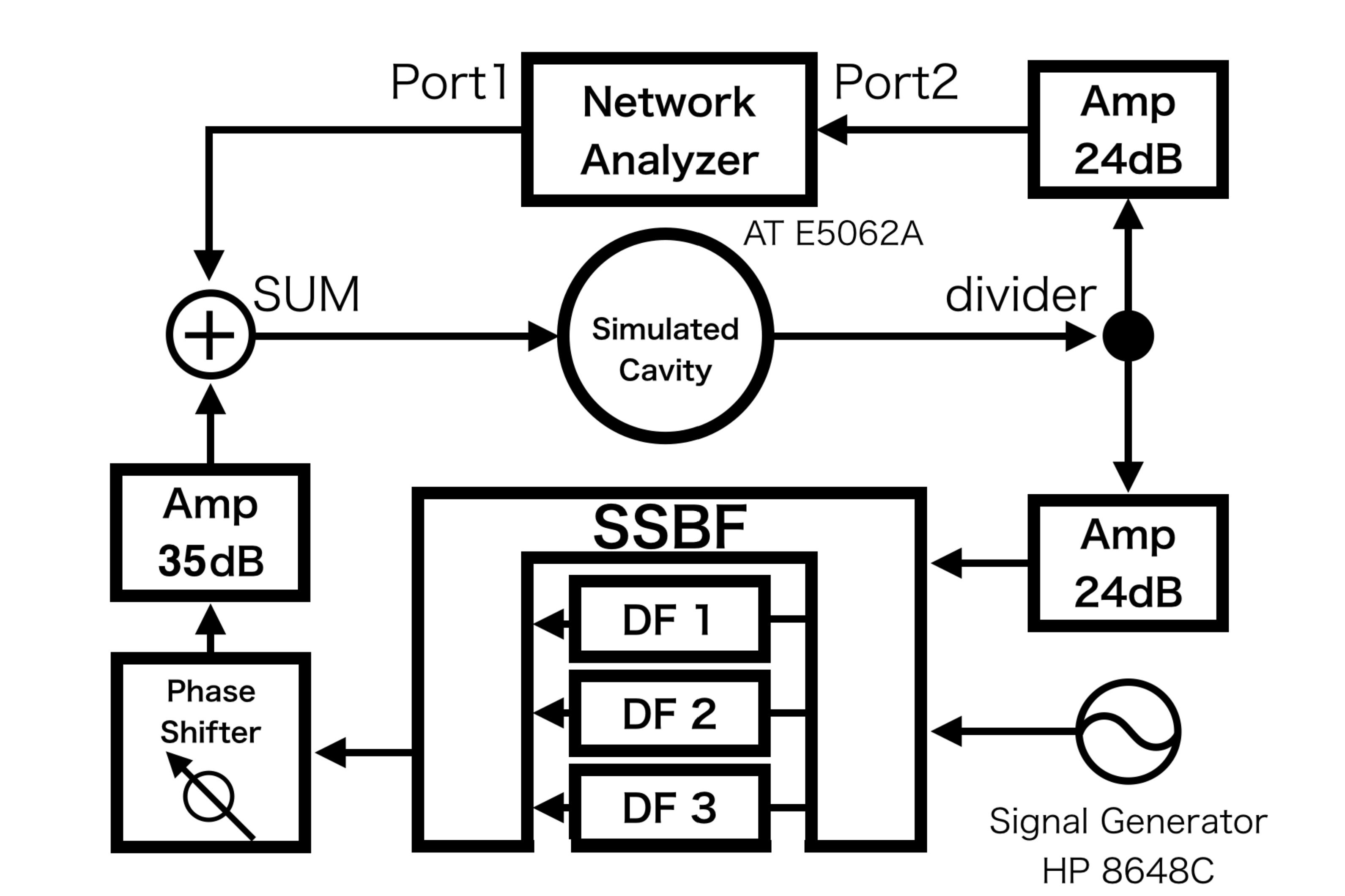}
		\includegraphics[width=\linewidth, bb = 0 0 453 348]{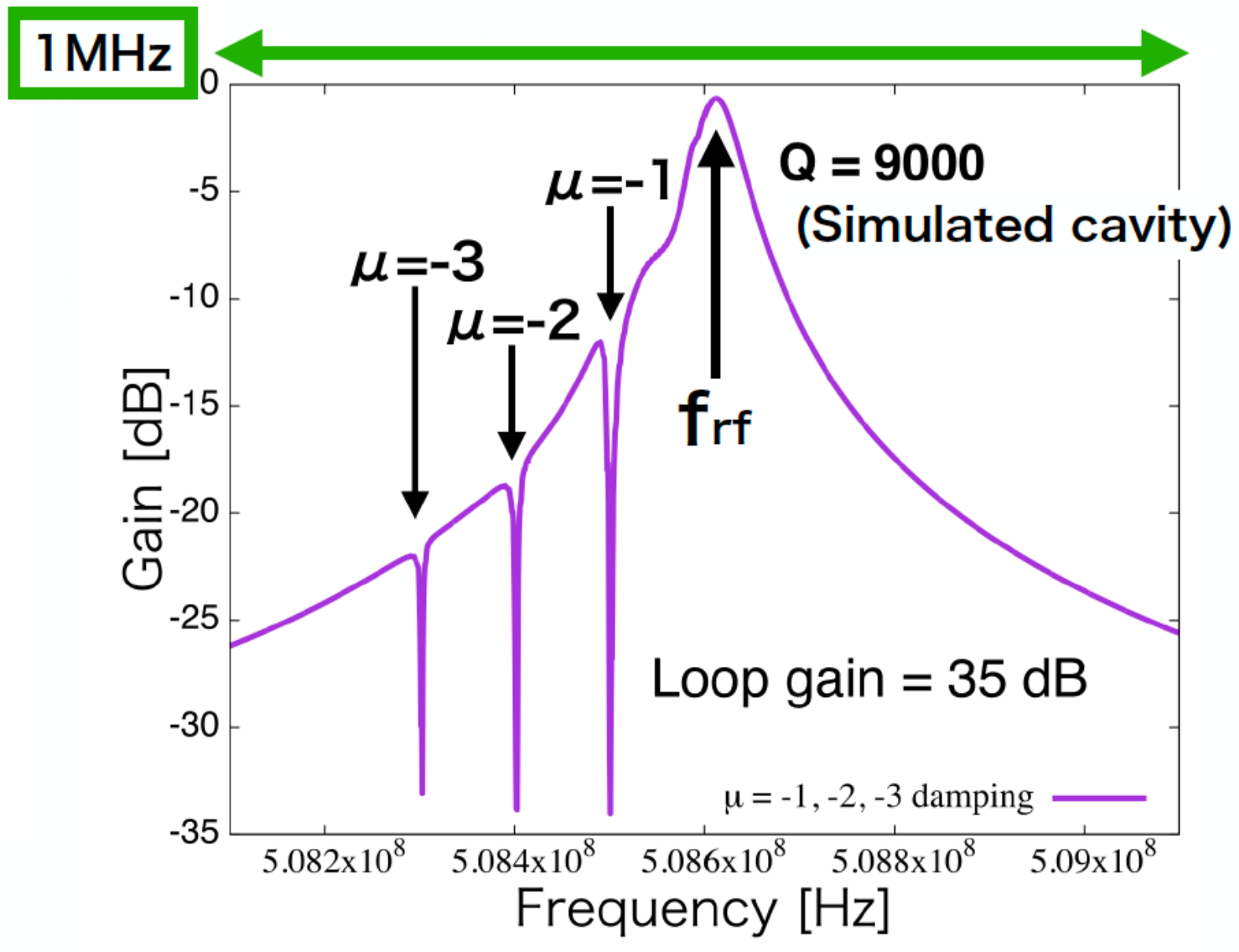}
		\caption{Block diagram for the FB loop evaluation (Top), and the damping characteristics for the FB loop (Bottom).}
		\label{fig:7}
	\end{figure}

	We have produced two SSBFs for LER and HER.
The evaluation result of SSBF performance is shown in Figure \ref{fig:6}.
Frequency characteristics of transmission measured by network analyzer are plotted for both of SSBFs in the figure.
The characteristics of the KEKB damper is also shown in the figure for comparison.
While the KEKB damper has a rejection of 40-dB for the stopband, a new damper gives 80-dB rejection.
With respect to the flatness of passband, a new damper has a very flat characteristic in an effective interval of frequency ($\mu$ = -1, -2, -3 mode excitation).
The performance of the new damper is improved obviously.
\\ \hspace{1em}
	Before practical use at beam commissioning, we have to make sure of LCBI damper performance.
	The nucleus of this damping instabilities method is attenuation of apparent impedance only at frequency that excites LCBI.
	Figure \ref{fig:7} shows the test-bench measurement of impedance damping with a simulated cavity(Q = 9000).
	The top figure is setup condition of FB evaluation, and the bottom figure is the result of characterization measurement which is the cavity impedance under FB worked.
	\\ \hspace{1em}
	The impedance is reduced only at $\mu$ = -1, -2, -3 modes depending on loop-gain.
	In the measurement, we optimized phases for a good characteristic by changing 5$^{\circ}$ or 10$^{\circ}$ from $\mu$ = -1 to $\mu$ = -3 in order.
	The desired result was obtained in FB loop performance for practical beam operations.
	At SuperKEKB Phase-2 commissioning, we will try to suppress beam oscillations by using new LCBI damper.

\section{Estimation of the Feedback effect for LCBI $\mu = -1$ mode}
\begin{figure}[!htb]
		\centering
		\includegraphics*[width=\linewidth, bb = 0 0 699 197]{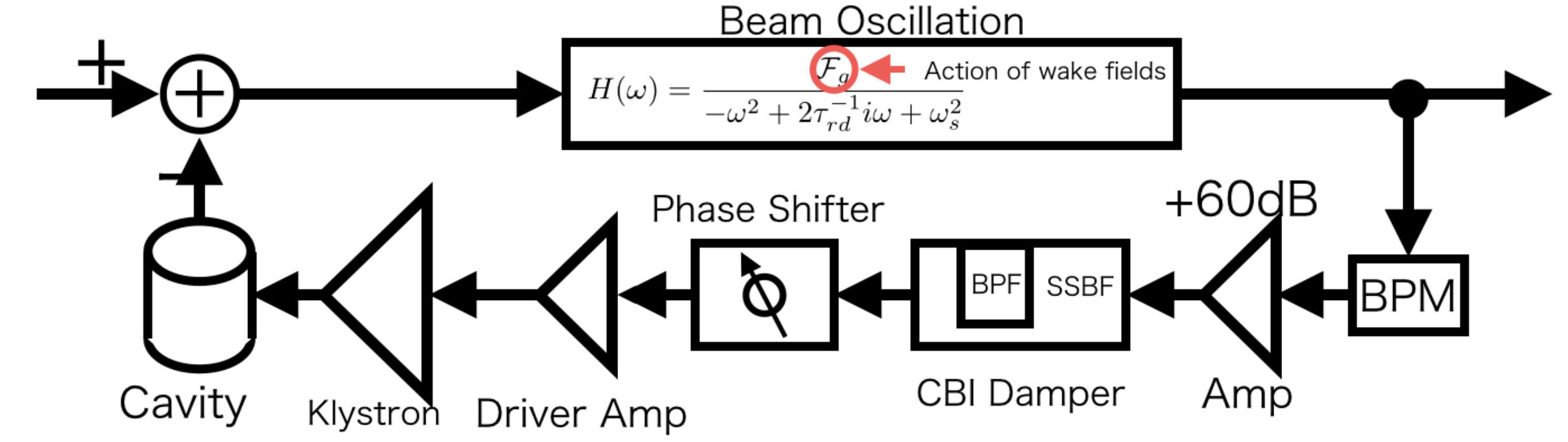}\\
		\includegraphics[width=\linewidth, bb = 0 0 846 594]{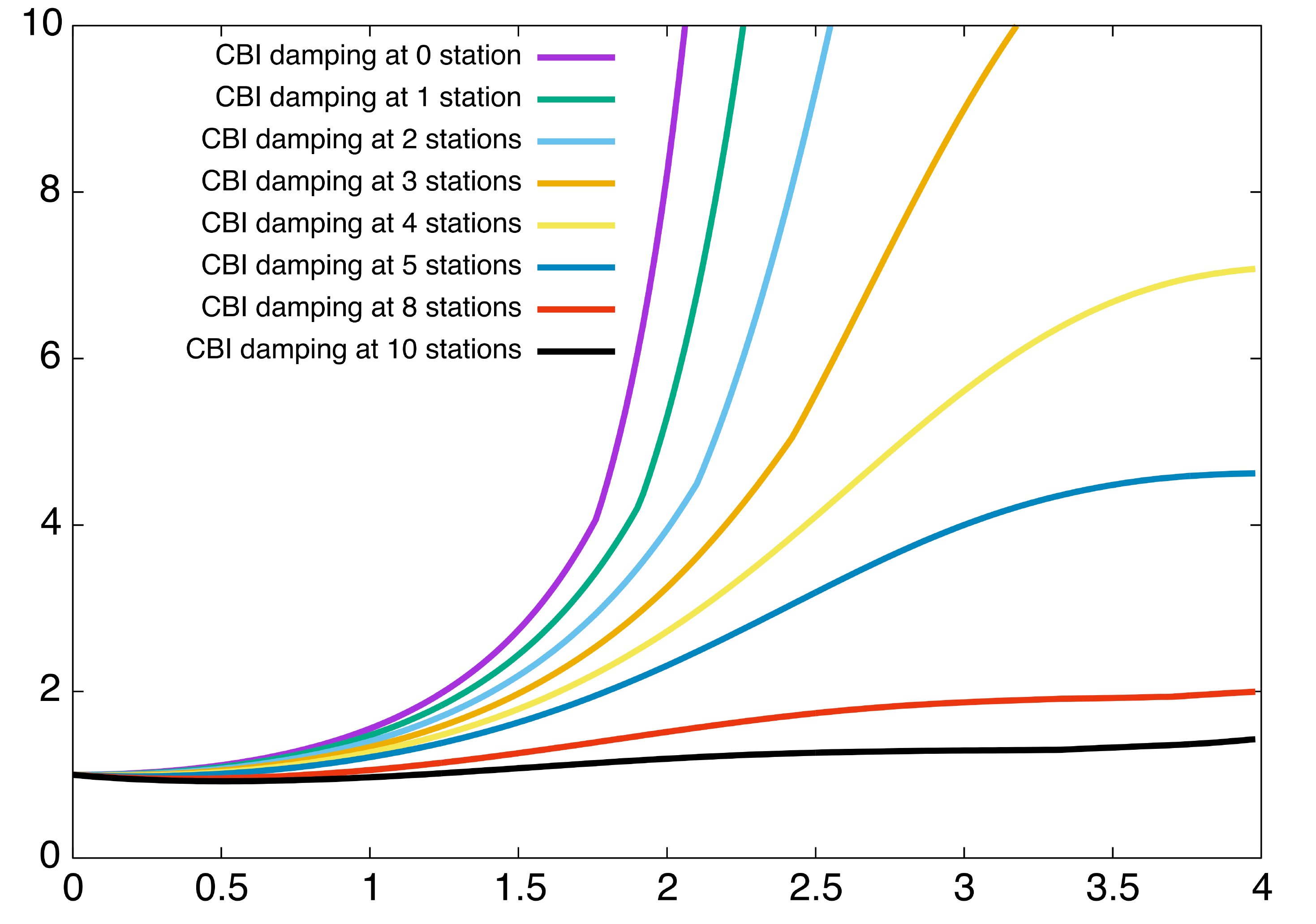}
		\caption{Estimation of the effectiveness of LCBI damping for the different numbers of the LCBI dampers applied to each RF station. Top: a block diagram for this simulation. Bottom: calculation results for each simulations.}
		\label{fig:8}
	\end{figure}

	We calculated RF FB to suppress multi-bunch oscillations,
	because the number of LCBI damper applied to cavities cannot be identified without evaluation of loop gain.
	The main idea is that bunches are kicked by wake fields and FB RF fields at cavities.
	The LCBI is an instability caused by the interaction between the impedance of the entire accelerator and the multi-bunch beam.
	If LCBI dampers are applied to all RF stations of storage ring, it is enough to supply the damping effect over the entire circumference of the ring.
	However it is necessary to suppress LCBI with the installation of LCBI dampers as few units as possible (for example, it is hard to install in stations of SC cavity because SC is more sensitive than NC).
	\\ \hspace{1em}
	This simulation is evaluation of the effect of difference of the number of RF stations which LCBI damper is applied to.
	The relation between pickup signals and wake fields level in a cavity is estimated from SuperKEKB Phase-1 commissioning measurement.
	Figure \ref{fig:8} shows estimation of the effectiveness of LCBI damping in beam-current change for the difference of the number of the LCBI damper applied to each RF station for LER (LCBI dampers and RF stations have a one-to-one relationship).
	In this figure, horizontal axis denotes beam current, and vertical axis is amplitude ratio of bunch oscillations to zero-current condition.  Initial gain of pickup signal is fixed to +60 dB (since intensity of pickup signals was so weak).
	\\ \hspace{1em}
	The gain control system may be required, so we plan to consider a new system including it after Phase-2 commissioning.
	As an estimation result, it was found to need some LCBI dampers for reach SuperKEKB design beam current.
	
\section{Preparation and the future plan}
\begin{figure}[!htb]
		\centering
		\includegraphics[width=0.94\linewidth, bb = 0 0 303 490]{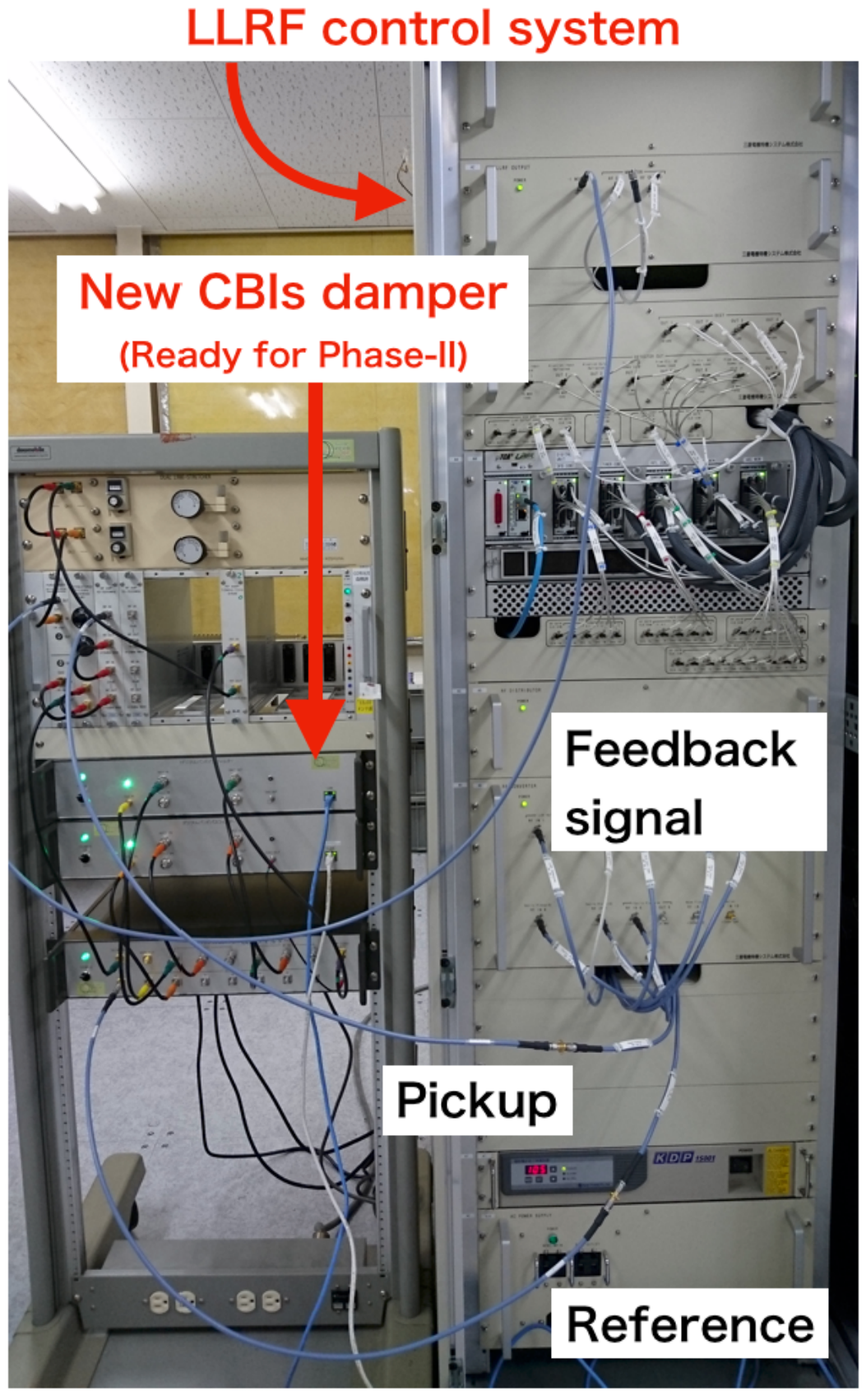}
		\caption{A picture of the preparation for using LCBI damper in Phase-2.}
		\label{fig:9}
	\end{figure}

\begin{figure}[!htb]
		\centering
		\includegraphics[width=\linewidth, bb = 0 0 762 310]{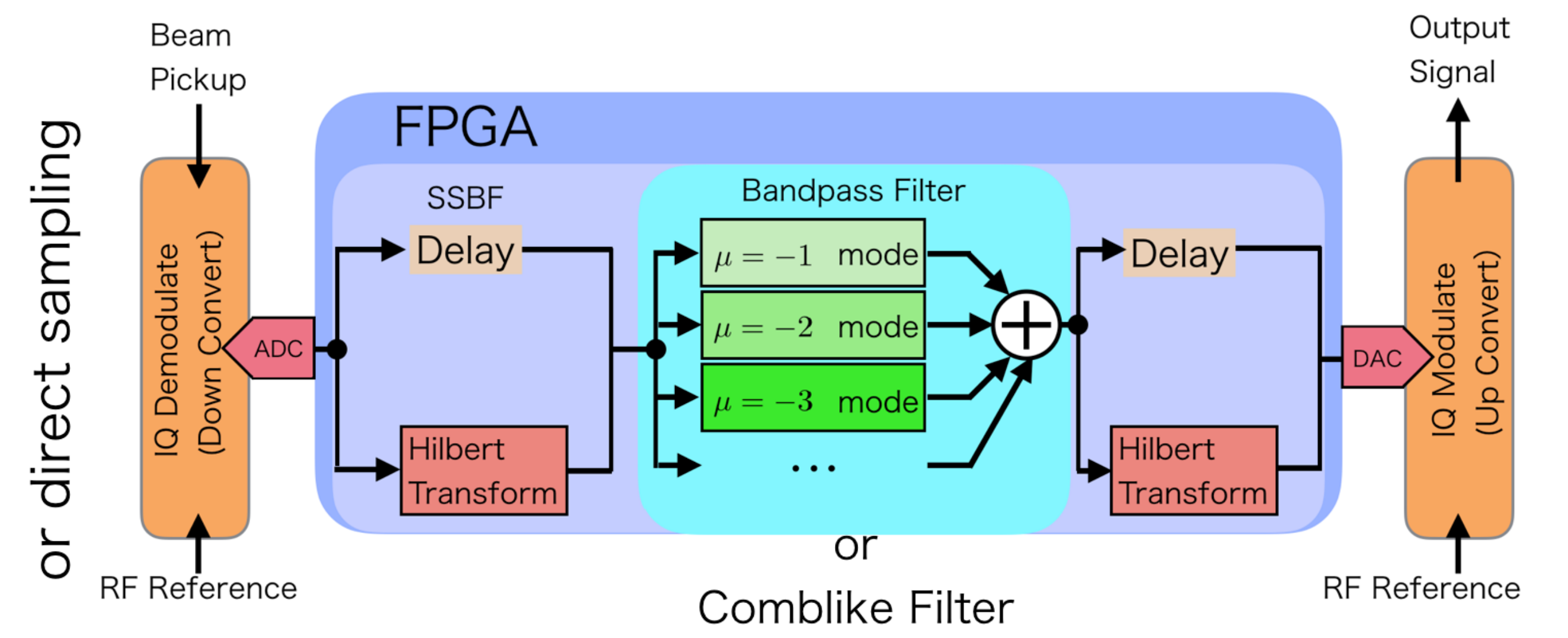}
		\caption{An illustration of design for improved LCBI damper.}
		\label{fig:10}
	\end{figure}

	The preparation for using LCBI damper in Phase-2 commissioning has already finished. Figure \ref{fig:9} is a picture of LCBI damper prepared for an RF station of NC.
	The damper is connected by three cables: for feedback, pickup, and reference from master oscillator. This pickup signal can be switched from cavity pickup to beam pickup.
	In KEKB operation and SuperKEKB Phase-1 commissioning, we used beam pickup signals, but we plan to use cavity pickup signals for getting larger gain in Phase-2 commissioning.
	Actual beam test in Phase-2 is also the test for confirming effectiveness of using cavity pickup signals.
\\ \hspace{1em}
For next step of this damper, we consider that all digitalized LCBI damper which is performed by FPGA. Figure \ref{fig:10} shows design of LCBI damper for the next version. Bandpass filters and main structure of system are kept from latest version, and SSBF is provided by Hilbert transformer. If it is possible, we will use direct sampling for analog-digital conversion. For SuperKEKB RF system, BPF is full digitalized damper is easier to produce additionally, so it is important for high luminosity accelerators.
	
\section{Conclusion}
	Developed new LCBI damper performance satisfies our required specifications.
	According to results of characteristic evaluation, the new SSBF provides an 80-dB reduction for stopbands, and passbands exhibit good flatness.
	SSBF and digital BPF was improved and their characteristics is very well.
	In the feedback loop characteristics, impedance damping is confirmed at the respective LCBI modes.
	In calculation of LCBI growth-rate (especially $\mu = -1$ mode), amplitude of bunch oscillation with beam current increase depends strongly on the number of the dampers applied.
	It is found that we need more LCBI dampers than one damper, but we still have not clarified the exact number of dampers required, so we will define how many dampers should be applied for the design beam current from beam test in Phase-2. 
	We expect that LCBI can suppress by new LCBI damper in SuperKEKB Phase-2 commissioning.

\end{document}